\documentclass[aps,prl,twocolumn,superscriptaddress,showpacs,nofootinbib,longbibliography]{revtex4-2}
\usepackage{graphicx}
\usepackage{dcolumn}
\usepackage{bm}
\usepackage{hyperref}
\usepackage{upgreek}
\usepackage{color}
\usepackage{soul}
\usepackage{epstopdf}
\usepackage{units}
\usepackage{longtable}
\usepackage{floatrow}
\usepackage{mhchem}

\begin{document}

\title{Molecular Beam Epitaxy of a Half-Heusler Topological Superconductor Candidate YPtBi}

\author{Jiwoong Kim}
\affiliation{Institute for Topological Insulators and Experimentelle Physik III, Universität Würzburg, 97074 Würzburg, Germany}
\affiliation{Max-Planck-Institut für Chemische Physik fester Stoffe, 01187 Dresden, Germany}
\author{Kajetan M. Fijalkowski}
\affiliation{Institute for Topological Insulators and Experimentelle Physik III, Universität Würzburg, 97074 Würzburg, Germany}
\author{Johannes Kleinlein}
\affiliation{Institute for Topological Insulators and Experimentelle Physik III, Universität Würzburg, 97074 Würzburg, Germany}
\author{Claus Schumacher}
\affiliation{Institute for Topological Insulators and Experimentelle Physik III, Universität Würzburg, 97074 Würzburg, Germany}
\author{Anastasios Markou}
\affiliation{Max-Planck-Institut für Chemische Physik fester Stoffe, 01187 Dresden, Germany}
\author{Charles Gould}
\affiliation{Institute for Topological Insulators and Experimentelle Physik III, Universität Würzburg, 97074 Würzburg, Germany}
\author{Steffen Schreyeck}
\affiliation{Institute for Topological Insulators and Experimentelle Physik III, Universität Würzburg, 97074 Würzburg, Germany}
\author{Claudia Felser}
\affiliation{Max-Planck-Institut für Chemische Physik fester Stoffe, 01187 Dresden, Germany}
\author{Laurens W. Molenkamp}
\affiliation{Institute for Topological Insulators and Experimentelle Physik III, Universität Würzburg, 97074 Würzburg, Germany}
\affiliation{Max-Planck-Institut für Chemische Physik fester Stoffe, 01187 Dresden, Germany}

\date{\today}

\begin{abstract}

The search for topological superconductivity has motivated investigations into materials that combine topological and superconducting properties. The half-Heusler compound YPtBi appears to be such a material, however experiments have thus far been limited to bulk single crystals, drastically limiting the scope of available experiments. This has made it impossible to investigate the potential topological nature of the superconductivity in this material. Experiments to access details about the superconducting state require sophisticated lithographic structures, typically based on thin films. Here we report on the establishment of high crystalline quality epitaxial thin films of YPtBi(111), grown using molecular beam epitaxy on Al$_{2}$O$_{3}$(0001) substrates. A robust superconducting state is observed, with both critical temperature and critical field consistent with that previously reported for bulk crystals. Moreover we find that AlO$_{x}$ capping sufficiently protects the sample surface from degradation to allow for proper lithography. Our results pave a path towards the development of advanced lithographic structures, that will allow the exploration of the potentially topological nature of superconductivity in YPtBi.

\end{abstract}
\maketitle

The emergence of topological states of matter \cite{Hasan2010} has opened up new possibilities in the context of quantum computation \cite{Kitaev2003,Qi2011}. One interesting idea involves the combination of topological and superconducting properties to create a so-called topological superconductor, with zero energy modes suitable for applications in quantum computing \cite{Fu2008,Qi2011}. One candidate family of materials expected to provide the conditions for the existence of such modes are half-Heusler compounds containing heavy elements \cite{Chadov2010}. Relativistic corrections in these materials lead to a band inversion with the fourfold degenerate \textit{p}-like ${\Gamma}_8$ band having higher energy than the twofold degenerate \textit{s}-like ${\Gamma}_6$ band \cite{Meinert2016}. A particularly promising member of this material class is YPtBi \cite{Chadov2010}. The \textit{j} = 3/2 character of the ${\Gamma}_8$ band emerging from the Bi \textit{p}-orbital states is predicted to lead to higher order pairing, and thus to a topological superconducting state \cite{Kim2018}. Previous experimental studies limited to bulk single crystals have shown the material to be superconducting below \textit{T$_{c}$} = 0.77 K, and to have a low hole-type carrier density of 2 ${\times}$ 10$^{18}$ cm$^{-3}$ \cite{Butch2011}. Moreover, the surface band structure was probed by angle-resolved photoemission spectroscopy, and confirmed to have non-trivial topological ordering \cite{Liu2016}. These results established that the material hosts both superconducting and topological properties, but there has been no evidence to date that the non-trivial topology leads to any exotic properties on the superconducting state.

In order to reveal any exotic nature of superconductivity it is typically necessary to examine lithographically patterned devices, where details of the superconducting wave function play a tangible role in the transport properties. Examples of such structures are superconducting quantum interference devices (SQUIDs) and Josephson junctions. The fabrication of such devices requires the establishment of advanced nanofabrication technology, typically based on thin film material, and is not really feasible on bulk crystals. The epitaxial growth of the half-Heusler compounds with  highly-ordered single crystal structure and well-defined interfaces can be produced using molecular beam epitaxy (MBE)\cite{Patel2016, Scheffler2020}. The present study takes two steps towards the realisation of precision patterned structures: 1) The demonstration of high quality epitaxial thin film growth of superconducting YPtBi using MBE, providing better growth control than in half-Heusler films grown by sputtering \cite{Shirokura2022, Bhardwaj2021} and 2) the demonstration of a capping technology based on AlO$_{x}$, in order to protect the thin film surface from degradation during fabrication.

%----------------------------------------------------------------------------FIG1

\begin{figure*}[!]
\includegraphics[width=\columnwidth]{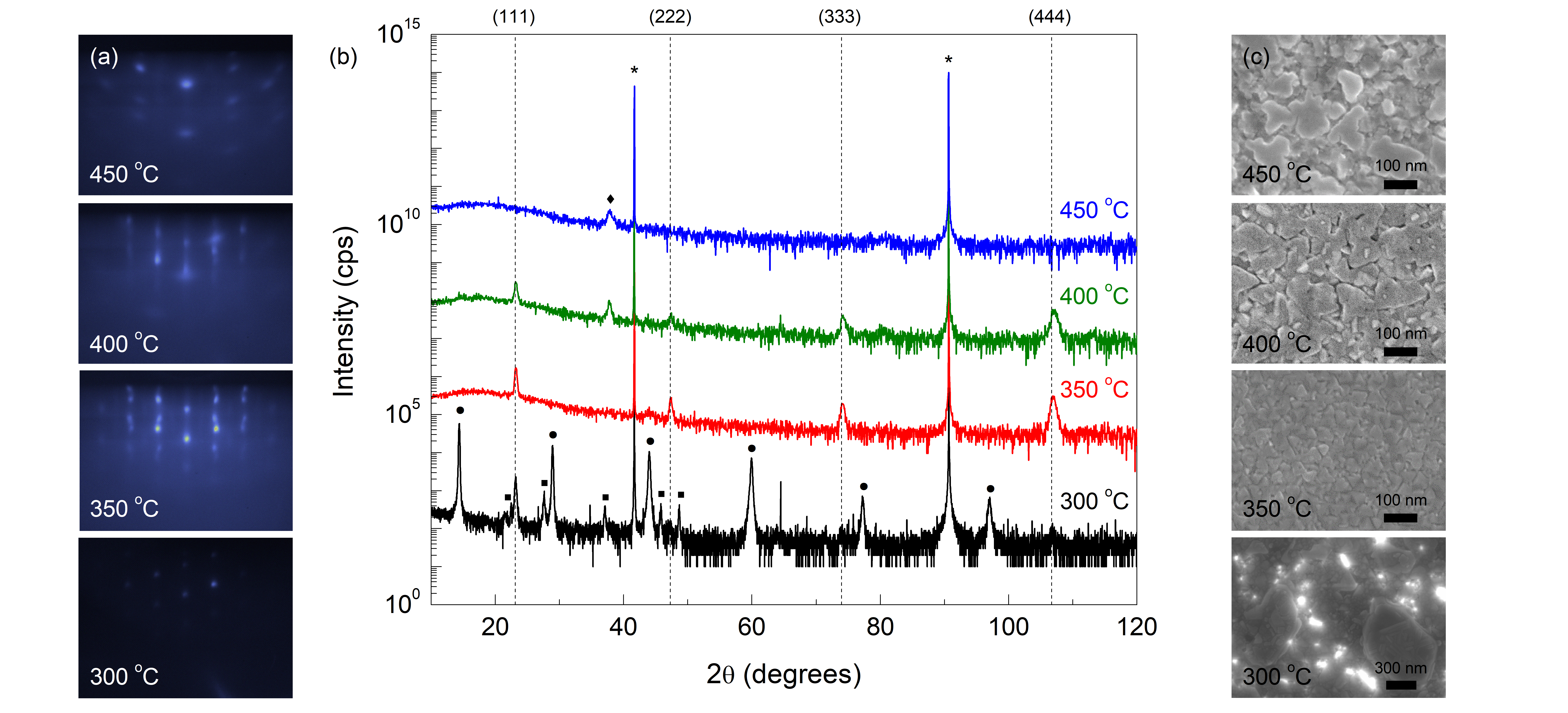}
\caption{Growth temperature dependence of the structural properties of YPtBi thin films. (a) RHEED patterns taken in-situ immediately after growth for each YPtBi thin film. The layers were grown with substrate temperatures of 300 $^{\circ}$C, 350 $^{\circ}$C, 400 $^{\circ}$C, and 450 $^{\circ}$C. (b) 2$\theta$-$\theta$ XRD diffractograms for each film. The expected positions of the (111) peaks of the YPtBi layer (\textit{F}$\bar{4}$3\textit{m}) are shown as dotted lines. The diffraction peaks of the Al$_{2}$O$_{3}$(0001) substrate, as well as those of monoclinic Bi, hexagonal Bi, and Pt$_{3}$Y are indicated by asterisks, circles, squares, and diamond, respectively. The curves are offset for clarity. (c) Scanning electron microscope images of the films.} 
\label{fig:Fig1}%
\end{figure*}
%----------------------------------------------------------------------------

Films of approximately 50 nm thickness are grown on Al$_{2}$O$_{3}$(0001) substrates that have a lattice mismatch of +1.2\% to the (111)-oriented YPtBi crystal structure. Prior to the growth, the substrate is annealed at 150 $^{\circ}$C for an hour to remove organic contaminants and water molecules. The three source materials; Y (4N purity), Pt (4N), and Bi (6N), are evaporated from elemental effusion cells. The beam equivalent pressure (BEP) of each element is monitored by an ionization gauge (Bayard-Alpert type) near the substrate position. The cell temperature for evaporating Pt is fixed at 1760 $^{\circ}$C providing a BEP of 1.6 ${\times}$ 10$^{-8}$ mbar, whereas the temperatures of the other cells are adjusted to tune the composition.

Figure 1 shows the structural and morphological properties of the thin films, which are affected by substrate temperature during growth. The Y BEP is fixed at 0.6 ${\times}$ 10$^{-8}$ mbar. The Bi:Pt BEP ratio is 2:1 for the film with substrate temperature of 300 $^{\circ}$C, and 4:1 for the remaining three films. During growth, the surface crystal structure is monitored in-situ using reflective high-energy electron diffraction (RHEED) (Fig. 1(a)). The electron beam was incident along the Al$_{2}$O$_{3}$[120] direction. The RHEED pattern of the samples grown at 300 $^{\circ}$C, 350 $^{\circ}$C, and 400 $^{\circ}$C are that of the YPtBi (111) surface plane, with the electron beam incident along the YPtBi[1$\bar{5}$4] direction. As the substrate temperature is increased from 350 $^{\circ}$C to 400 $^{\circ}$C, the pattern changes from spotty to streaky as the increased kinetic energy of the atoms improves the surface structure and helps to make it flat. The RHEED pattern from the 450 $^{\circ}$C sample, on the other hand, shows the (111) surface of Pt$_{3}$Y, consistent with the X-ray diffraction (XRD) analysis. 

Figure 1(b) shows 2$\theta$-$\theta$ XRD diffractograms of YPtBi films grown at various substrate temperatures. The 300 $^{\circ}$C sample shows diffraction peaks from an elementary (100)-oriented monoclinic Bi (C12/\textit{m}1 \cite{Brugger1967}) phase as well as a polycrystalline hexagonal Bi (\textit{R$\bar{3}$m:H} \cite{Cucka1962}) phase. These are indicated by black circles and squares, respectively. The samples grown at the three higher temperatures do not show any signs of elemental Bi. This is attributed to a sufficiently high thermal energy to desorb Bi atoms from the surface. For the samples grown at temperatures from 300 $^{\circ}$C to 400 $^{\circ}$C, the peak at 23.1$^{\circ}$ stems from the (111) plane of half-Heusler YPtBi, and demonstrates the crystalline growth of YPtBi on the Al$_{2}$O$_{3}$(0001) plane with (111)-preferred orientation. The XRD curve of the sample grown at 400 $^{\circ}$C also has a diffraction peak associated with Pt$_{3}$Y(111), suggesting insufficient Bi flux. This Bi-deficient phase is the only one present in the XRD diffractogram for the sample grown at 450 $^{\circ}$C.

Figure 1(c) shows scanning electron microscope (SEM) images of the surface of each film. Clusters are visible on the surface of the sample grown at 300 $^{\circ}$C, which are determined by energy dispersive X-ray spectroscopy to be elemental Bi. In contrast, the other three films show only triangular-shaped grains arising from hexagonal YPtBi(111). The size of the triangular grains increases with substrate temperature. Surface of the film grown at 450 $^{\circ}$C is considerably rougher, with arbitrarily shaped grains.

%----------------------------------------------------------------------------FIG2
\begin{figure}[!]
\includegraphics[width=\columnwidth]{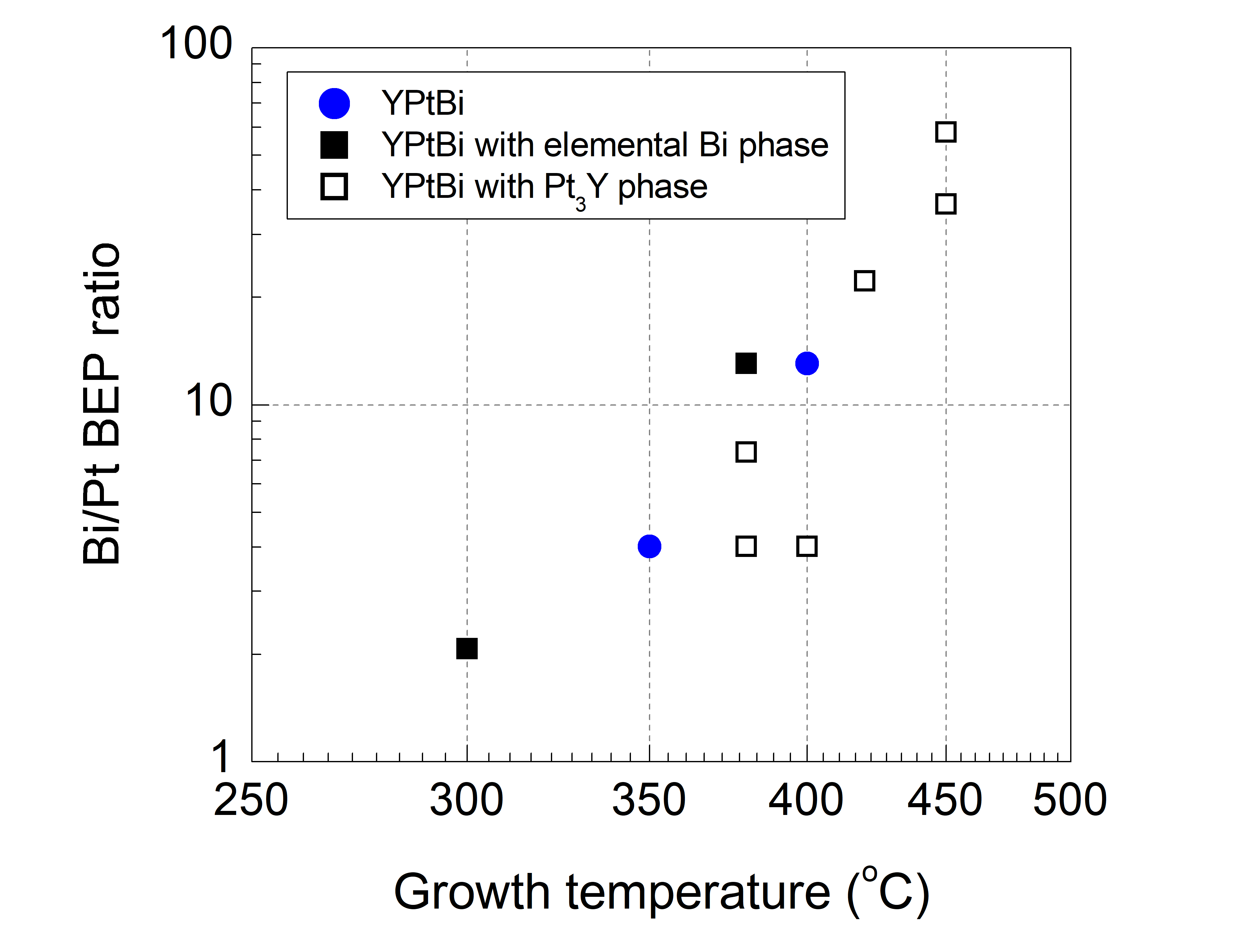}
\caption{
 Phase space for YPtBi growth based on the XRD analysis. The YPtBi(111) thin films with elemental Bi phase and Pt$_{3}$Y phase are depicted as black filled squares and empty squares, respectively. Two YPtBi(111) films with no detectable secondary phases are shown as blue circles. 
}%
\label{fig:Fig2}%
\end{figure}
%----------------------------------------------------------------------------

Next we investigate the growth regime of YPtBi(111) films in terms of relative ratio of Bi to Pt BEPs. In all cases, the growth is performed under excess Bi BEP, as Bi incorporation is self-limited due to its volatility \cite{Shourov2020, Patel2016}. We define the ratio of the BEP of Bi with respect to that of Pt, which is fixed at 1.6 ${\times}$ 10$^{-8}$ mbar. The Y BEP is fixed at 0.6 ${\times}$ 10$^{-8}$ mbar throughout the series. Figure 2 gives the phase diagram with the various phases identified by XRD analysis. Thin films showing only diffraction peaks of YPtBi(111) without any secondary phases are depicted as blue circles. The black filled and empty squares indicate samples showing the YPtBi phase with an elemental Bi phase and a Pt$_{3}$Y phase, respectively, in the XRD diffractograms. The two samples where no secondary phases were detected roughly identify the growth regime of pure YPtBi(111). This diagram also indicates that an increase in growth temperature necessitates an increase in Bi flux in order to retain a pure YPtBi phase.

%----------------------------------------------------------------------------FIG3
\begin{figure}[!]
\includegraphics[width=\columnwidth]{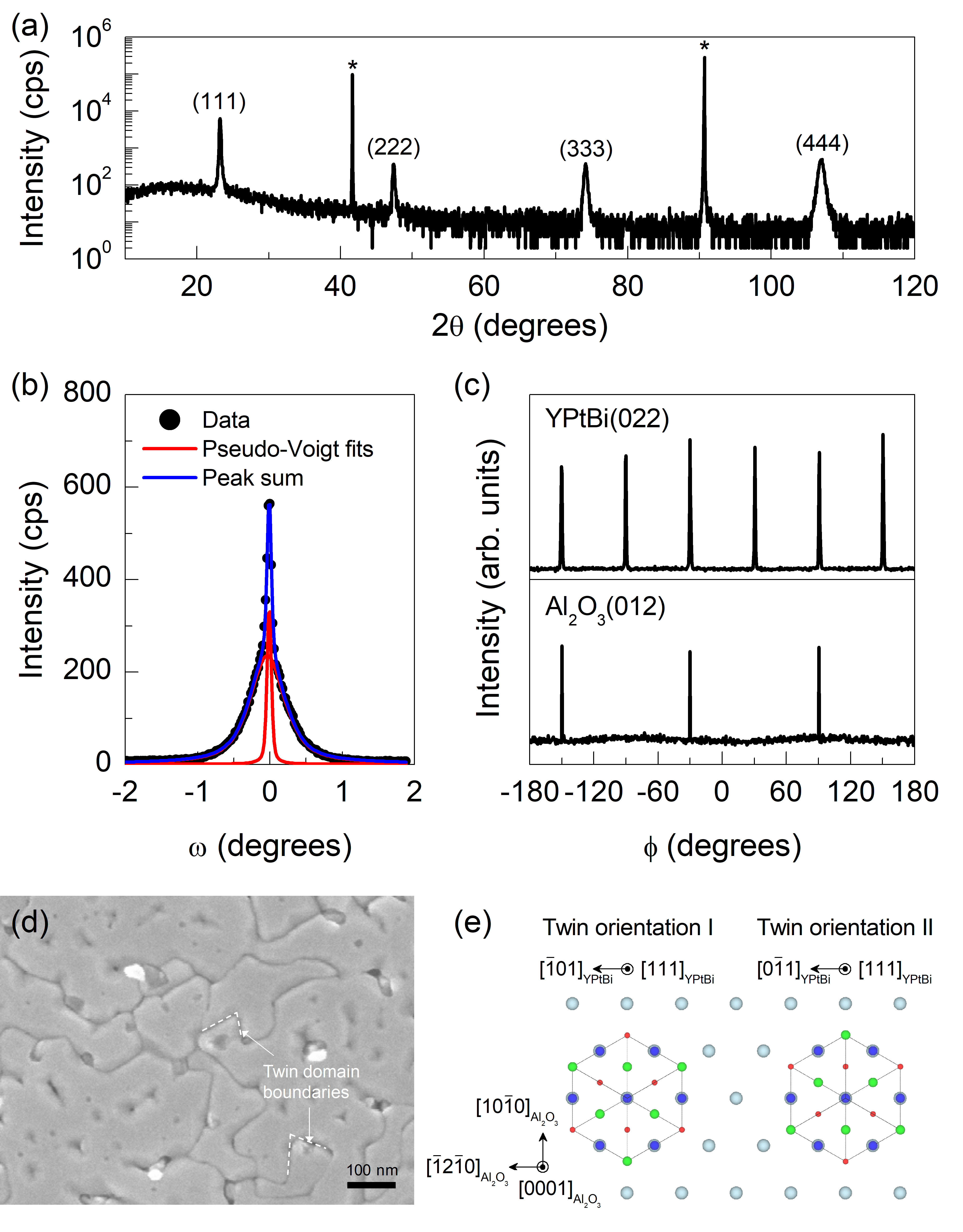}
\caption{
Crystal structure and surface morphology of a YPtBi(111) thin film grown under optimum growth conditions, as confirmed by transport experiments. (a) 2$\theta$-$\theta$ XRD diffractogram of the film. The indices label diffraction peaks of YPtBi. Diffraction peaks from the Al$_{2}$O$_{3}$(0001) substrate are marked with asterisks. (b) Rocking curve of the YPtBi(111) diffraction peak. The peak is deconvoluted into two pseudo-Voigt curves. (c) Azimuthal rotation $\phi$-scans of the asymmetric (022) plane of YPtBi and the (012) plane of Al$_{2}$O$_{3}$. (d) SEM image of the film's surface morphology. Distinct grain boundaries are highlighted with white dashed lines, illustrating the 60$^{\circ}$ angles due to twinned growth of (111)-oriented YPtBi crystals. (e) Schematic of the atomic configurations of (111)-oriented YPtBi crystals on the Al-terminated Al$_{2}$O$_{3}$(0001) surface.
}%
\label{fig:Fig3}%
\end{figure}
%----------------------------------------------------------------------------

The crystalline quality and surface morphology was further improved by tuning the Y BEP. Figure 3 summarizes the structural properties of the YPtBi thin film which was grown at the Y BEP of 1.1 ${\times}$ 10$^{-8}$ mbar and a substrate temperature of 400 $^{\circ}$C with 13:1 Bi:Pt BEP ratio. Figure 3(a) shows a 2$\theta$-$\theta$ diffractogram for YPtBi(111)${\Vert}$Al$_{2}$O$_{3}$(0001). The full width at half maximum (FWHM) of the YPtBi(111) diffraction peak is 0.21$^{\circ}$, which is narrower than the film grown at the lower Y flux (0.6 ${\times}$ 10$^{-8}$ mbar) (0.48$^{\circ}$). Figure 3(b) shows the rocking curve of the YPtBi(111) diffraction peak. The peak is deconvoluted into two pseudo-Voigt curves with FWHMs of 0.07$^{\circ}$ and 0.58$^{\circ}$. The narrower part of the peak confirms the presence of highly oriented YPtBi crystallites. The broader feature implies rotational disorder in the epitaxial film \cite{Miceli1995}. The lattice mismatch (1.2\%) between the YPtBi(111) and Al$_{2}$O$_{3}$(0001) planes may generate misfit dislocations that lead to small displacements in the epitaxial YPtBi layer. This disorder gives short-range correlations resulting in the diffusive feature. The FWHM of the broad peak is nevertheless also narrower than the peak for the film grown at lower Y flux (0.99$^{\circ}$), indicating that disordered crystallites are reduced.

The in-plane crystalline symmetry and orientation was examined by an azimuthal rotation $\phi$-scan of an asymmetric plane of both the YPtBi epilayer and the substrate (Fig. 3(c)). The six-fold YPtBi(022) diffraction peaks reveal twinned YPtBi(111) crystal domains. The half-Heulser YPtBi crystal structure has a 3-fold rotational symmetry with respect to the (111) axis, and a comparison between the two $\phi$-scans indicates that the YPtBi crystal domains are oriented towards either YPtBi[$\bar{1}$01]${\Vert}$Al$_{2}$O$_{3}$[$\bar{1}$2$\bar{1}$0] or YPtBi[0$\bar{1}$1]${\Vert}$Al$_{2}$O$_{3}$[$\bar{1}$2$\bar{1}$0] on the Al$_{2}$O$_{3}$(0001) substrate, as sketched in Fig. 3(e). The SEM image in Fig. 3(d) shows the surface morphology of the film. The average grain size is larger than that of the sample shown in Fig. 1(c). The triangular-shaped boundaries between grains may result from the stacking of (111)-oriented YPtBi crystals on the lattice mismatched substrate (1.2\%). 

%----------------------------------------------------------------------------FIG4
\begin{figure*}[!]
\includegraphics[width=\columnwidth]{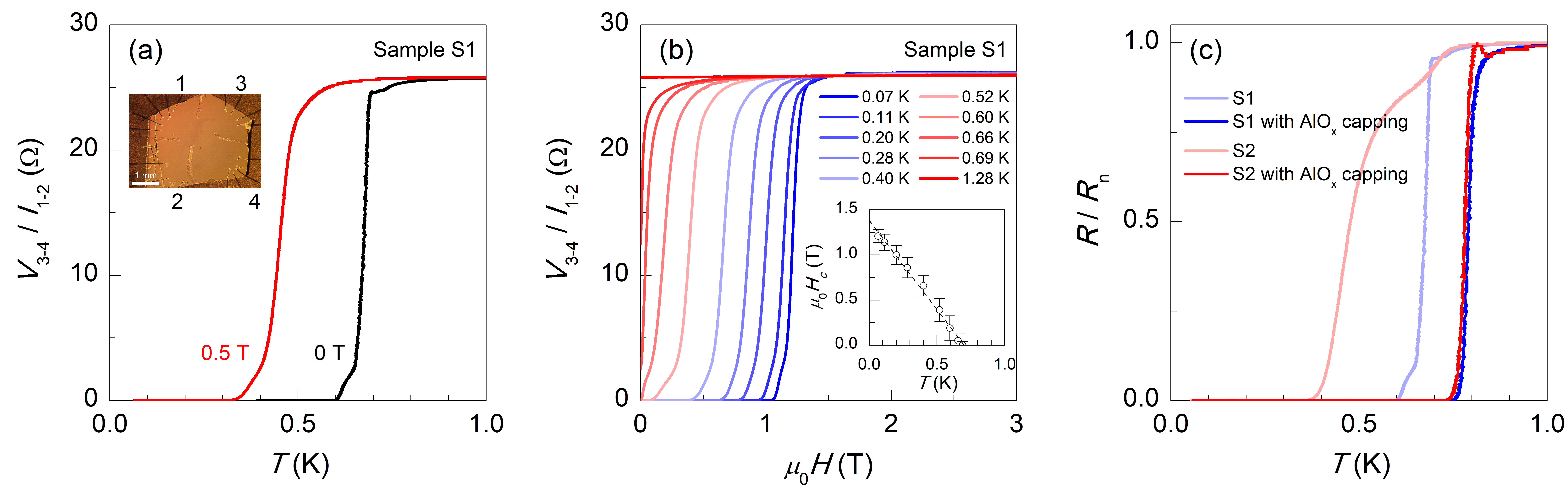}
\caption{
(a) Temperature dependence (cool down curves) of the four-terminal resistance of the sample S1, with an out of plane external magnetic field of 0 T or 0.5 T. Au wires bonded on the sample surface, and used for the 4-terminal measurement are indicated as numbers on the inset image. (b) Magnetic field-dependence of resistance of the YPtBi(111) film at various temperatures. As a benchmark for the critical field \textit{H$_{c}$} we take the magnetic field where the resistance is half of the normal state resistance, and plotted it in the inset. The zero temperature value of \textit{H$_{c}$} is estimated to be 1.4 T based on the linear extrapolation. This value is compatible with the one previously reported for bulk YPtBi (1.5 T \cite{Butch2011}). (c) Influence of AlO$_{x}$ capping on the YPtBi(111) thin film transport properties. A 5 nm-thick AlO$_{x}$ layer was deposited using atomic layer deposition (ALD), without exposing the YPtBi film surface to air. Layers S1 and S2 were grown under nominally identical growth conditions to the optimized layer presented in Fig. 3. The 4-terminal resistance (\textit{R$=V_{3-4}/I_{1-2}$}) for each sample is normalized to the normal state resistance (\textit{R}$_{n}$) at 1 K. 
}%
\label{fig:Fig4}%
\end{figure*}
%----------------------------------------------------------------------------

Transport measurements are performed on films grown under nominally the same conditions as the optimized layer in Fig. 3. The inset of Fig. 4(a) shows an optical microscope image of one of the samples used in the measurements. The surface was scratched with a diamond tip into the “clover-shaped” geometry to minimize the influence of the contacts \cite{Pauw1958} and to separate the clamp shadow area, located at the left hand side, from the device. The Au wires bonded to the sample and used for the electrical measurements are labeled as the numbers in the inset image. The superconducting phase transition is probed by collecting the 4-terminal resistance ($V_{3-4}/I_{1-2}$) while cooling the sample in a magnetic field of either 0 T or 0.5 T, applied perpendicular to plane of the sample. The obtained critical temperature (\textit{T$_{c}$}) is 0.67 K and 0.48 K at 0 T and 0.5 T, respectively. The zero field \textit{T$_{c}$} value is lower than that previously observed in the bulk sample (0.77 K) \cite{Butch2011}. Hall resistance measurement ($V_{3-2}/I_{1-4}$) taken at fields above the critical field show linear dependence up to 8 T, with \textit{p}-type carrier transport, and a carrier concentration of 1.7 ${\times}$ 10$^{20}$ cm$^{-3}$ at 1.3 K, obtained utilizing the van der Pauw geometry. This is some two orders of magnitude higher than that previously reported for bulk YPtBi (2 ${\times}$ 10$^{18}$ cm$^{-3}$ \cite{Butch2011}). The carrier mobility is 64 cm$^{2}$/Vs. 

The temperature dependence of the critical field (\textit{H$_{c}$}) was obtained by measuring the sample magneto-resistance at various temperatures (Fig. 4(b)). The value of \textit{H$_{c}$} is taken at the magnetic field where the resistance is half of the normal state resistance, with the error bars defined as values corresponding to 10\% to 90\% of the normal state resistance. The \textit{H$_{c}$} values are plotted as a function of temperature in the inset of Fig. 4(b). \textit{H$_{c}$} has a linear dependence on temperature, unlike conventional superconductors where a parabolic trend is expected \cite{Tinkham2004}. From extrapolation, the zero-temperature critical field is estimated to be about 1.4 T. This value is comparable to the Pauli paramagnetic limit (\textit{${\mu}$}$_{0}$\textit{H$_{p}$} ${\approx}$ 1.86\textit{T$_{c}$} = 1.3 T) \cite{Orlando1979}, and to the previously reported bulk crystal value (1.5 T \cite{Butch2011}). The superconducting coherence length (in the out-of-plane direction) is estimated to be ${\xi}$ = 15 nm by using the Ginzburg-Landau formula \cite{Tinkham2004} which is much shorter than the film thickness (50 nm). 

While the above results are promising, the obtained critical temperature is noticeable below that of bulk crystals, quite possibly because of degradation at the exposed surface. To address this issue, we applied an AlO$_{x}$ capping technology \cite{Logan2016} to investigate the protection of the surface of the films.

Figure 4(c) shows the temperature dependence of the four-terminal longitudinal resistance measurements on four samples, with the values normalized at 1 K. Films labeled S1 and S2 are grown under nominally the same growth conditions as the one shown in Fig. 3. These films are extracted from the growth chamber and cleaved in N$_{2}$ environment. Sample pieces are then inserted into an atomic layer deposition (ALD) chamber using a glove bag, such that the transfer between chambers happens without the surface being exposed to air. A 5 nm-thick AlO$_{x}$ layer is deposited on a cleaved piece from each of  S1 and S2. The \textit{T$_{c}$} of the uncapped samples S1 and S2 is 0.67 K and 0.48 K, respectively, while the capped samples show 0.79 K and 0.78 K, in a good agreement with bulk YPtBi \cite{Butch2011}. This clearly demonstrates that the \textit{T$_{c}$} is reduced when the YPtBi thin film is exposed to air, and that AlO$_{x}$ capping adequately protect the surface.

In summary, we have investigated the crystal structure and transport properties of superconducting epitaxial YPtBi thin films. (111)-oriented half-Heusler YPtBi thin films are grown on Al$_{2}$O$_{3}$(0001) substrate by MBE, and optimal growth conditions are identified. The resulting film properties are found to suffer from surface degradation. This issue is counteracted by adding an AlO$_{x}$ layer on the films without exposing them to air. Films with the thus protected surface have a critical temperature matching that of bulk samples. These thin films provide a platform on which lithography can be developed. This combination will enable a variety of experiments that require microscopic patterned samples, an important step towards potential verification of topological nature of superconducting state in YPtBi.

\begin{acknowledgments}

We gratefully acknowledge the financial support of the Free State of Bavaria (the Institute for Topological Insulators), Würzburg-Dresden Cluster of Excellence on Complexity and Topology in Quantum Matter CT.QMAT (EXC 2147, 39085490), the Deutsche Forschungsgemeinschaft (DFG, German Research Foundation)-Projektnummer (392228380), and the ERC Advanced Grant No. (742068) “TOP-MAT”

\end{acknowledgments}

%\bibliography{Refs.bib}


\end{document}